\documentclass[aps,twocolumn,showpacs,superscriptaddress,groupedaddress]{revtex4}
\usepackage{color,graphicx,calc,url}
\usepackage{dcolumn}
\usepackage{bm}
\usepackage{amssymb}
\usepackage{amsmath}
\usepackage{wasysym}
\usepackage{epstopdf}

\begin{document}
\title{Impact of strain and morphology on magnetic properties of Fe$_3$O$_4$/NiO bilayers grown on Nb:SrTiO$_3$(001) and MgO(001)}

\author{O.~Kuschel}
\affiliation{Department of Physics and Center of Physics and Chemistry of New Materials, Osnabr{\"u}ck University, 49076 Osnabr{\"u}ck, Germany}

\author{N.~Path\'{e}}
\affiliation{Department of Physics and Center of Physics and Chemistry of New Materials, Osnabr{\"u}ck University, 49076 Osnabr{\"u}ck, Germany}

\author{T.~Schemme}
\affiliation{Department of Physics and Center of Physics and Chemistry of New Materials, Osnabr{\"u}ck University, 49076 Osnabr{\"u}ck, Germany}

\author{K.~Ruwisch}
\affiliation{Department of Physics and Center of Physics and Chemistry of New Materials, Osnabr{\"u}ck University, 49076 Osnabr{\"u}ck, Germany}

\author{J.~Rodewald}
\affiliation{Department of Physics and Center of Physics and Chemistry of New Materials, Osnabr{\"u}ck University, 49076 Osnabr{\"u}ck, Germany}

\author{R.~Bu\ss}
\affiliation{Department of Physics and Center of Physics and Chemistry of New Materials, Osnabr{\"u}ck University, 49076 Osnabr{\"u}ck, Germany}


\author{F.~Bertram}
\affiliation{DESY, Photon Science, 22607 Hamburg, Germany}

\author{T.~Kuschel}
\affiliation{Center for Spinelectronic Materials and Devices, Department of Physics, Bielefeld University, 33615 Bielefeld, Germany}

\author{K.~Kuepper}
\affiliation{Department of Physics and Center of Physics and Chemistry of New Materials, Osnabr{\"u}ck University, 49076 Osnabr{\"u}ck, Germany}

\author{J.~Wollschl\"ager}\email{jwollsch@uos.de}
\affiliation{Department of Physics and Center of Physics and Chemistry of New Materials, Osnabr{\"u}ck University, 49076 Osnabr{\"u}ck, Germany}


\date{\today}

\keywords{}

\begin{abstract}
We present a comparative study on the morphology and structural as well as magnetic properties of crystalline Fe$_3$O$_4$/NiO bilayers grown on both MgO(001) and SrTiO$_3$(001) substrates by reactive molecular beam epitaxy. 
These structures are investigated by means of x-ray photoelectron spectroscopy, low energy electron diffraction, x-ray reflectivity and diffraction as well as vibrating sample magnetometry.
While the lattice mismatch of NiO grown on MgO(001) is only 0.8\,\%, it is exposed to a lateral lattice mismatch of -6.9\,\% if grown on SrTiO$_3$. 
In the case of Fe$_3$O$_4$, the misfit strain on MgO(001) and SrTiO$_3$(001) amounts to 0.3\,\% and -7.5\,\%, respectively.
To clarify the relaxation process of the bilayer system, the film thicknesses of the magnetite and nickel oxide films have been varied between 5 and 20\,nm.
While NiO films are well ordered on both substrates, Fe$_3$O$_4$ films grown on NiO/SrTiO$_3$ exhibit a higher surface roughness as well as lower structural ordering compared to films grown on NiO/MgO.
Further, NiO films grow pseudomorphic in the investigated thickness range on MgO substrates without any indication of relaxation, whereas on SrTiO$_3$ the NiO films show strong strain relaxation. 
Fe$_3$O$_4$ films exhibit also strong relaxation even for films of 5\,nm thickness on both NiO/MgO as well as on NiO/SrTiO$_3$.
The magnetite layers on both substrates show a fourfold magnetic in-plane anisotropy with magnetic easy axes pointing in $\left\langle100\right\rangle$ directions.
The coercive field is strongly enhanced for magnetite grown on NiO/SrTiO$_3$ due to higher density of structural defects, compared to magnetite grown on NiO/MgO. 


\end{abstract}

\maketitle

\section{Introduction}

Transition metal oxides are one of the most interesting material classes providing a huge variety of structural, magnetic, and electronic properties ranging from metallic to insulating, from ferro- to antiferromagnetic, as well as ferroelectric states \cite{transmetox}. Especially, thin magnetite films (Fe$_3$O$_4$) attracted intensive research interest in the last decade in the field of spintronics \cite{spintronic} and spin caloritronics \cite{spincal,moussy}. 
Due to their anticipated half-metallic behavior with complete spin polarization at the Fermi level \cite{halfmet} and high (bulk) Curie temperature of 858\,K \cite{BookofIronOxide}, thin magnetite films are promising candidates for room temperature spintronic devices such as highly spin-polarized electrodes for magnetic tunneling junctions \cite{mtj, highTMR, LucaMTJMgO} or spin injectors \cite{spininjector}. 
Furthermore, multilayers of magnetite and platinum show huge thermoelectric effects \cite{ramos16} based on the recently observed spin Seebeck effect in magnetite \cite{ramos13} pushing the development of more efficient thermoelectric nanodevices \cite{SSE}.

Magnetite crystallizes in the inverse spinel structure with a lattice constant of 8.3963\,\AA~\cite{BookofIronOxide} at 300\,K. At $\sim$\,120\,K it undergoes a metal-insulator transition (Verwey transition) \cite{verwey} accompanied by a change from cubic to monoclinic crystal symmetry \cite{MagMonoclin}. The reduction of the crystal symmetry leads to a spontaneous ferroelectric polarization and, thus, to multiferroicity \cite{ferroelctric, multiferroic}.

In order to control the relative magnetization alignment in magnetic tunnel junctions, exchange bias effects induced by additional antiferromagnetic layers are commonly used \cite{FirstExchangeCoupling}. 
In case of Fe$_3$O$_4$ tunnel junctions, the antiferromagnetic NiO is a good candidate due to its small lattice mismatch of only 0.5\,\% and a high N\'{e}el temperature of 523\,K \cite{NiONeel}.

Nickel oxide is an insulating material with a high thermal stability. 
It crystallizes in a rock salt structure with a lattice constant of 4.1769\,\AA~\cite{NiOPoisson} at 300\,K.
Recently it was shown that NiO can act as a spin current amplifier in spin Seebeck experiments, and additionally be a spin current generator when a thermal gradient is applied \cite{SpinCurrentAFM, SSEAFM, HashPoisson, Holanda2017}, making NiO a key material for thermoelectric devices.
Further, latest studies report on a temperature dependent sign change in the spin hall magnetoresistance for nickel oxide on ferromagnetic insulator and, thus, a possibility to use it as a spin filter \cite{Geert2017, Hou2017}.

Previous works \cite{Berti1,MagMomMag,StrainMagMgO,bilayerExBias,bilayersMgO,bilayersMgO2,SchemmeBilayer} have focused on characterization of magnetite and nickel oxide films grown on MgO substrates because of the small lattice mismatch of 0.3\,\% and 0.8\,\%, respectively. 
However, it has been demonstrated that the electronic and magnetic properties of magnetite films can be modified using SrTiO$_3$ substrates \cite{MagSTO, MagAxisMagSTO,Liu2017} in spite of the large lattice mismatch of -7.5\,\%. 
For instance, one advantage using SrTiO$_3$ substrates is the possibility of doping and, thus, a tunable conductivity providing either an insulating or metallic substrate which could be used as bottom electrode in capacitor-like structures \cite{multiferroic}. 
Furthermore, Fe$_3$O$_4$/NiO bilayers grown on SrTiO$_3$ can be used to synthesize Ni$_x$Fe$_{3-x}$O$_4$ thin films by thermally induced interdiffusion with tunable magnetic and electric properties \cite{Kuschel1}.

Up to now, most studies concerning NiO films on SrTiO$_3$ are limited to a coarse analysis of the growth \cite{NiOSTO,NiOSTO2}, while a thorough structural characterisation is seldom reported \cite{Zhang2017}.
In the case of Fe$_3$O$_4$/NiO bilayers on both substrates, there is a number of works on electronic structure, interfacial coupling and magnetic characterization \cite{MagPropBilayer,Kupper,Fe3O4NiOinterface,ElStructBilayer} whereas to the best of our knowledge, there are no detailed structural studies for these bilayers on SrTiO$_3$.
However, the magnetic and transport characteristics of such films are sensitive to structural variations, number of defects or stoichiometric deviations and could be affected by the strain between film and substrate \cite{Liu2017}. 
Therefore, in this work a comprehensive structural characterization of Fe$_3$O$_4$/NiO bilayers of different thicknesses grown on Nb-doped SrTiO$_3$(001) and for comparison on MgO(001) is presented. 
Additionally, these results are correlated to magnetic properties, e.g., magnetocrystalline anisotropy.

Directly after deposition, the stoichiometry in the near surface region and the surface structure of each layer was determined \textit{in situ} using x-ray photoelectron spectroscopy (XPS) and low energy electron diffraction (LEED), respectively. 
The bulk structure was investigated \textit{ex situ} by x-ray reflectivity (XRR) and synchrotron radiation x\,-\,ray diffraction (SR-XRD) measurements and analyzed within the kinematic diffraction theory. 
Further, angle dependent hysteresis loops are measured via vibrating sample magnetometry (VSM).
 
\section{Experimental Details}

Preparation and \textit{in situ} characterization of the thin oxide films were carried out in an interconnected ultrahigh\,-\,vacuum (UHV) system at a base pressure of 10$^{-8}$\,mbar in the deposition chamber and 10$^{-10}$\,mbar in the analysis chamber. 
Epitaxial Fe$_3$O$_4$/NiO ultra thin bilayer systems with thicknesses between 5\,nm and 20\,nm were grown via reactive molecular beam epitaxy (RMBE) on 0.05\,\% Nb-doped SrTiO$_3$(001) or on MgO(001) single crystalline substrates. 
Prior to deposition, the substrates were annealed at 400\,$^\circ$C in 1$\times$10$^{-4}$\,mbar O$_2$ atmosphere for 1\,h in order to remove carbon contamination and get well-defined surfaces. 
Subsequently, nickel oxide and magnetite films were deposited by thermal evaporation from pure metal rods in 1$\times$10$^{-5}$\,mbar and 5$\times$10$^{-6}$\,mbar oxygen atmosphere, respectively. 
Deposition was performed at 250\,$^\circ$C substrate temperature using deposition rates of 0.01\,nm/s for nickel oxide films and 0.25\,nm/s for magnetite films as controlled by a quartz microbalance adjacent to the evaporation source. 
The resulting film thicknesses were determined later on \textit{ex situ} by XRR. 
Crystal surface quality and near surface stoichiometry were controlled \textit{in situ} after each preparation step by LEED and XPS using an Al K$_\alpha$ (h$\nu\,=\,1486.6$\,eV) radiation source and a Phoibos HSA 150 hemispherical analyzer. 

\begin{figure}
\centering
\includegraphics[width=0.47\textwidth]{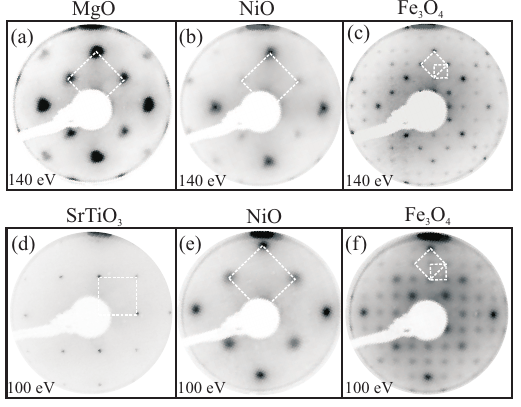}
\caption{LEED pattern recorded at 140\,eV for (a) pure MgO(001) surface, (b) 11.9\,nm NiO film on MgO(001) and (c) 21.5\,nm Fe$_3$O$_4$ on NiO/MgO(001). The LEED pattern taken at 100\,eV of a pure SrTiO$_3$ surface, a 10.4\,nm NiO film on SrTiO$_3$(001) and 20.7\,nm Fe$_3$O$_4$ on NiO/SrTiO$_3$(001) are depicted in (d), (e) and (f), respectively. The larger white square indicates (1$\times$1) structure of the reciprocal unit cell of the respective surface while the smaller white square in (c) and (f) indicates the ($\sqrt{2}\times\sqrt{2})R45^{\circ}$ superstructure unit cell of magnetite.}
\label{LEED}
\end{figure}

After transport under ambient conditions XRR and XRD experiments were carried out \textit{ex situ} for structural characterization of the films. 
XRR measurements were performed in $\theta$\,-\,$2\theta$ geometry using a lab based diffractometer (Philips X’Pert Pro) equipped with a Cu K$_\alpha$ anode. 
An in-house developed fitting tool based on the Parratt algorithm \cite{ParratXRR} using N\'{e}vot-Croce \cite{nevot} roughness profiles was applied for the analysis of the XRR curves. 
For XRD synchrotron based radiation sources at the MaXLab beamline I811 (MaXLab, Lund, Sweden) and at the Swiss Light Source beamline X04SA (Paul Scherrer Institute, Villigen, Switzerland) were used. 
Both beamlines are equipped with (2S\,+\,3D) type diffractometers and Pilatus pixel area detectors for data collection.
The XRD data were recorded in $\theta$\,-\,$2\theta$ geometry at an energy of 12.4\,keV and analyzed within the kinematic diffraction theory \cite{KinDifTheo} that is implemented in our in-house developed fitting tool.

In addition, magnetization curves were measured at room temperature for several in-plane directions of the samples by varying the magnetic field $\mu_0$$H$ between \mbox{-300\,mT} and +300\,mT, using a VSM (Lakeshore, Model~7407). 
The magnetization loops were corrected by subtracting the diamagnetic contribution from the substrates.
 
\section{Results}

\subsection{LEED\,/\,XPS}
Figures \ref{LEED}(a) and \ref{LEED}(d) present the LEED patterns of the cleaned MgO(001) and SrTiO$_3$(001) surfaces, respectively.
All as prepared NiO and Fe$_3$O$_4$ films show similar LEED patterns on the respective substrate for all investigated thicknesses ranging from 5\,nm to 20\,nm.
Thus, only patterns of a $\sim$20\,nm Fe$_3$O$_4$ and a $\sim$10\,nm NiO film on MgO and SrTiO$_3$ are shown exemplarily in Fig. \ref{LEED}. 
The intensity variations in all recorded patterns are due to dynamical scattering for electron diffraction and will not be considered further.
Instead we focus on the symmetry of the diffracted pattern and the sharpness of the diffraction spots. 

Clear (1$\times$1) structures corresponding to the square unit cells of MgO(001) and SrTiO$_3$(001) surfaces can be seen (cf.~Fig.~\ref{LEED}(a) and (d)). 
Due to the rocksalt structure of MgO the reciprocal unit vectors of the MgO(001) surface point in [110] and [$\bar{1}10$] directions forming a quadratic reciprocal unit cell. 
The reciprocal unit vectors of the (001) surface of the perovskite SrTiO$_3$, however, point in [100] and [010] directions forming a quadratic unit cell, as well. 
Consequently, the reciprocal surface unit vectors of MgO(001) are $\sim\sqrt{2}$ times larger than those of SrTiO$_3$(001). 
 
In diffraction patterns a random arrangement of point defects leads to an increased background while line defects as, e.g., domain boundaries result in a broadening of the diffraction spots \cite{henzler}.
To obtain not only qualitative but also quantitative information on the defect density the full width of half maximum (FWHM) of the diffraction spots was determined at 140\,eV taking into account the instrumental broadening of the LEED instrument.  

SrTiO$_3$ pattern exhibits sharp and intense diffraction spots.
Analysis of the FWHM of the (11) diffraction peaks yields a line defect density of (0.11$\pm$0.02)\,nm$^{-2}$. 
In contrast the spots of the MgO substrate are broadened due to charging effects.
Thus, it was not possible to determine a value for the defect density of the substrate here.
The diffuse background is quite low in both patterns pointing to clean surfaces and negligible point defects. 
Additional, XPS measurements of both substrates show no carbon contamination indicating chemically clean surfaces. 

After deposition of NiO the LEED patterns also exhibit a (1$\times$1) structure related to the square symmetry of the NiO(001) surface for both substrates (cf. Fig. \ref{LEED}(b) and (e)). 
As mentioned above, due to the rocksalt structure, the reciprocal unit vectors of NiO(001) surface point in [110] and [$\bar{1}10$] directions and are consequently $\sim\sqrt{2}$ times larger than the surface unit cell of SrTiO$_3$ in reciprocal space.
Due to the very similar lattice constants of NiO(001) and MgO(001) the diffraction spots are located at almost identical positions. 
A broadening of the diffraction spots compared to the pattern of the SrTiO$_3$ substrate is clearly visible indicating an increase of the defect density.
Analyzing the FWHM of the (10) surface diffraction spots, we obtain densities of line defects of (0.8$\pm$0.1)\,nm$^{-2}$ and (1.1$\pm$0.2)\,nm$^{-2}$ for the NiO/MgO and NiO/SrTiO$_3$, respectively.
The slightly larger broadening of the diffraction spots for NiO/SrTiO$_3$ compared to the diffraction spots of the NiO/MgO surface can be related to the formation of more structural defects, e.g., domain boundaries, induced by the higher lattice misfit of NiO(001) on SrTiO$_3$(001).
Additionally, both patterns show a negligible background intensity of the NiO(001) surface pointing to a small amount of point defects.
 
\begin{figure}
\centering
\includegraphics[width=0.43\textwidth]{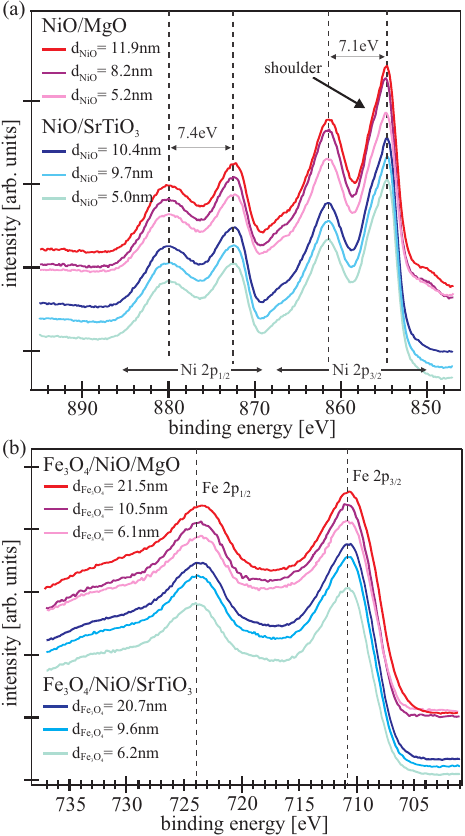}
\caption{X\,-\,ray photoelectron spectra of (a) Ni\,2p region for the as prepared NiO films on MgO(001) and SrTiO$_3$. (b) Fe\,2p region for the as prepared Fe$_3$O$_4$ films on NiO/MgO(001) and NiO/SrTiO$_3$.}
\label{XPS}
\end{figure}

\begin{figure*}[t]
\centering
\includegraphics[width=1\textwidth]{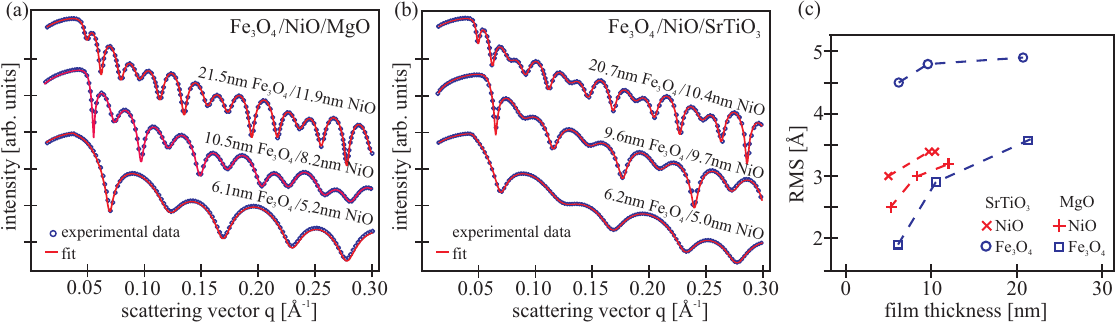}
\caption{XRR measurements and the calculated intensities of the bilayers on (a) MgO and (b) SrTiO$_3$ substrates. (c) Fe$_3$O$_4$ surface and Fe$_3$O$_4$/NiO interface roughnesses obtained from the XRR measurements.}
\label{XRR}
\end{figure*}

The LEED images of Fe$_3$O$_4$ obtained after deposition on NiO/MgO(001) and NiO/SrTiO$_3$(001) show similar diffraction patterns with a square symmetry (cf.~Fig.~\ref{LEED}(c) and (f)). 
Clear diffraction spots with half peak distance compared to the NiO(001) surface indicate approximately doubled lattice constant in real space due to the almost doubled cubic lattice constant of Fe$_3$O$_4$ compared to the other oxides used here. 
Furthermore, an additional ($\sqrt{2}\times\sqrt{2})R45^{\circ}$ superstructure appears which is characteristic for a well-ordered magnetite surface \cite{pentcheva,korecki,and97,Bliem2014}. 
This superstructure is not observed for maghemite (Fe$_2$O$_3$) which has a very similar surface lattice constant.
Therefore, we assume a formation of well-ordered stoichiometric magnetite films. 
However, the diffraction spots of the magnetite film grown on NiO/MgO are sharper than for the growth on NiO/SrTiO$_3$ indicating a better ordering and less domain boundaries.
For the density of line defects of the Fe$_3$O$_4$ films values of (1.3$\pm$0.2)\,nm$^{-2}$ and (0.14$\pm$0.02)\,nm$^{-2}$ are obtained for the growth on NiO/SrTiO$_3$(001) and NiO/MgO(001), respectively, analyzing the FWHM of the (20) surface diffraction spots. 

In summary, the LEED patterns of the Fe$_3$O$_4$/NiO bilayer systems confirm a crystalline cube on cube growth of both NiO and Fe$_3$O$_4$ films on MgO(001) as well as on SrTiO$_3$(001).
The films grown on MgO substrates exhibit a higher crystalline quality and less surface defects compared to the bilayers grown on SrTiO$_3$.

XPS measurements were made directly after deposition of the films to determine the stoichiometry and the valence state of the cation species.
Figure \ref{XPS}(a) shows the XP spectra of the Ni\,2p region after deposition of nickel oxide and before deposition of iron oxide. 
All spectra of the Ni\,2p core level reveal Ni\,2p$_{3/2}$ and Ni\,2p$_{1/2}$ peaks at binding energies of 854.6\,eV and 872.5\,eV, respectively, and two intense satellite structures at about 7\,eV higher binding energies. 
Since these values agree well with the binding energies reported in literature for a Ni$^{2+}$ valence state in NiO stoichiometry \cite{NiOSatellites, NiOenergies} we assume that the oxide films are stoichiometric and have negligible amounts of point defects as, e.g., oxygen vacancies. 
Additionally, there is a shoulder $\sim$1.5 eV above the Ni\,2p$_{3/2}$ peak, which is reported to be typical for NiO \cite{uhlen,NiOshoulder}. 
Thus, the shape of all spectra is comparable to that of NiO bulk crystal \cite{XPSNiO,NiOenergies,XPSNiO2}.

In Figure \ref{XPS}(b), the Fe\,2p photoelectron spectra of the iron oxide films as prepared on top of the NiO films are presented. 
From the position and shape of the Fe\,2p peaks one can get information about the iron oxidation state and the stoichiometry. 
All recorded spectra exhibit the same shape with main peaks located at binding energies of 710.6\,eV and 723.6\,eV for Fe\,2p$_{3/2}$ and Fe\,2p$_{1/2}$, respectively. 
These binding energies of the core levels correspond to values of Fe$_3$O$_4$ well-known from literature \cite{Yamashita}. 
Additionally, in contrast to w\"ustite (FeO) and maghemite (Fe$_2$O$_3$), no apparent charge transfer satellites can be observed between the two main peaks due to their overlap \cite{Yamashita,Fuji}. 
Consequently, the shape and binding energies of the Fe\,2p spectra confirm a mixed Fe$^{2+}$/Fe$^{3+}$ valence and point to a Fe$_3$O$_4$ stoichiometry for all prepared iron oxide films.
Thus, both XPS and LEED measurements demonstrate that the bilayer structures on both kind of substrates consist of crystalline stoichiometric NiO and Fe$_3$O$_4$ films. 



\subsection{XRR\,/\,XRD}
\begin{figure*}[tb]
\centering
\includegraphics[width=1\textwidth]{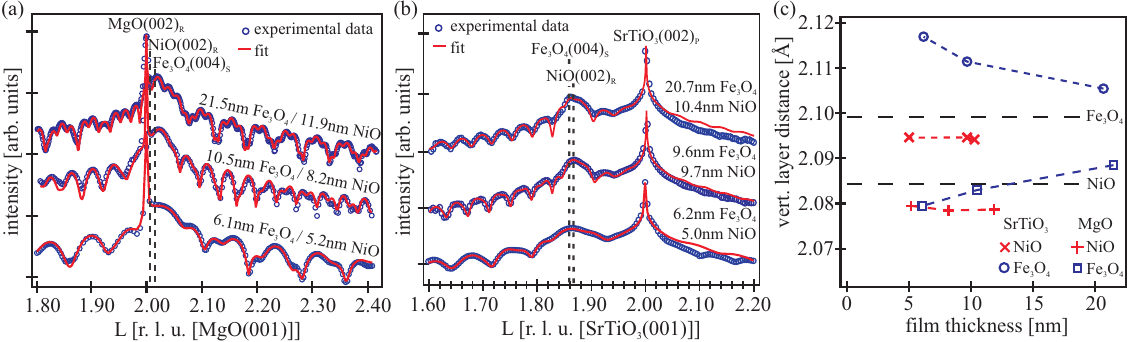}
\caption{XRD measurement along the (00$L$) CTR (a) of the Fe$_3$O$_4$/NiO/MgO samples and (b) of the Fe$_3$O$_4$/NiO bilayers on SrTiO$_3$. In red the calculated intensity distribution using the kinematic approximation is shown. (c) Vertical layer distance of nickel oxide and magnetite grown on MgO(001) and SrTiO$_3$(001) dependent on the film thickness. The dashed lines denote the fully relaxed bulk values of magnetite and nickel oxide.}
\label{XRD}
\end{figure*}
XRR and XRD experiments were performed \textit{ex situ} to determine the structural parameters of the bilayers, e.g., film thicknesses and vertical lattice distances.
Figures \ref{XRR}(a) and \ref{XRR}(b) show the measured reflectivity curves and the corresponding calculated reflectivity curves after optimizing the structural parameters. 
In addition, the obtained thicknesses of all studied bilayers are presented. 
For all samples clear intensity oscillations with beating effects are visible indicating double layer structures and flat homogenous films with small interface and surface roughness.

The applied calculation model consists of a layer of iron oxide on top of a nickel oxide layer on MgO or SrTiO$_3$ substrate.
All fitted curves agree excellently with the experimental data using literature values for the dispersion $\delta_{\textrm{Fe$_3$O$_4$}}$\,=\,$1.53\times 10^{-5}$ and $\delta_{\textrm{NiO}}$\,=\,$1.89\times 10^{-5}$ \cite{henke}. 
This indicates a small defect density, e.g., oxygen vacancies which is in accordance with the XPS results.

Additionally, the roughnesses of the films were determined and are presented in Fig. \ref{XRR}(c). 
Here, all films feature an increase of the surface and interface roughness with increasing film thickness.
This effect can be attributed to kinetic roughening of the films during growth and to the progressing relaxation process \cite{KinRoughning}.
The nickel oxide films exhibit similar roughnesses of $\sigma_{\textrm{NiO}}$\,=\,2.5\,-\,3.5\,\AA\ on both substrates with small increase for thicker films.
The roughness of the Fe$_3$O$_4$ on NiO/MgO increases more drastically, while the magnetite films deposited on NiO/SrTiO$_3$ show nearly constant roughness with initially almost doubled values compared to the magnetite films on NiO/MgO. 
This behavior is likely caused by high lattice misfit and the resulting relaxation process.
It is in accordance with the broadened diffraction spots of the Fe$_3$O$_4$ films on NiO/SrTiO$_3$ observed in the LEED pattern (cf.~Fig.~\ref{LEED}). 

Figures \ref{XRD}(a) and \ref{XRD}(b) present the SR-XRD measurements of the (00$L$) crystal truncation rod (CTR) compared to intensities calculated by kinematic diffraction theory of the Fe$_3$O$_4$/NiO bilayers on MgO(001) and SrTiO$_3$(001), respectively. 
Here, the bulk nomenclature of the reciprocal space was used, where $L$\,=\,$c\,K$$_{\perp}$/(2$\,\pi$) in reciprocal lattice units (r.l.u.) denotes the vertical scattering vector $K_{\perp}$ scaled to the Bragg condition 2\,$\pi$/$c$ ($c_{\textrm{MgO}}$\,=\,4.2117\,\AA, $c_{\textrm{SrTiO}_3}$\,=\,3.905\,\AA). 
The diffraction data reveal an epitaxial (001)-oriented growth of NiO and Fe$_3$O$_4$ on both substrates.
Due to the almost doubled lattice constant of magnetite compared to both MgO and NiO and the resulting lateral tensile strain, the (004)$_S$ spinel bulk reflection is located at higher $L$ values compared to MgO and close to the (002)$_R$ bulk reflection of a rock salt structure. 
On SrTiO$_3$, both nickel oxide and magnetite exhibit a large lattice misfit and are laterally compressively strained. 
Thus, the (004)$_S$ reflection of magnetite and (002)$_R$ reflection of NiO are at lower $L$ values compared to SrTiO$_3$ and well separated from the (002)$_P$ perovskite reflection of SrTiO$_3$.
Here, the indexes $R$, $S$ and $P$ indicate bulk indexing for rock salt, spinel and perovskite type, respectively.

For all bilayers grown on MgO, the measurements show a sharp peak at $L$\,=\,2 originating from the diffraction at the MgO substrate lattice (cf.~Fig.~\ref{XRD}(a)). 
Additionally, broad rather intense features located at $L$\,$\sim$\,2.02 accompanied by strong Laue oscillations are visible due to finite thickness of the iron and nickel oxide films. 
The well\,-\,pronounced intensity oscillations with two superposed partial oscillations clearly show a periodicity of two layers of different thickness indicating a high crystalline ordering and homogenous thicknesses of both films, magnetite and nickel oxide.
This is in accordance with the results seen in the XRR measurements.

In the case of bilayers grown on SrTiO$_3$, the (00$L$) rod also shows a sharp substrate peak at $L$\,=\,2 and Laue oscillations due to crystalline magnetite and nickel oxide films (cf.~Fig.~\ref{XRD}(b)).
Here, the Bragg peaks originating from the iron and nickel oxide are located at $L$\,$\sim$\,1.86 and broadened due to the finite film thicknesses. 
On closer inspection, the Laue oscillations show also a periodicity of two layers whereby the damping of the oscillation originating from the magnetite surface increases with increasing magnetite thickness due to increasing roughness (cf.~Fig.~\ref{XRR}(c)).
This result agrees well with LEED and XRR results shown above.

Due to the small lattice mismatch between Fe$_3$O$_4$ and NiO a separation of the Bragg peaks originating from the respective film is not visible by eye. 
Complete data analysis using kinematic diffraction theory was performed to obtain the vertical layer distance of the respective oxide film. 
Within the calculation the atomic form factors of oxygen, nickel and iron atoms arranged in a bulk structure are kept constant while the vertical size of the unit cell is varied. 
Interface roughness is modeled with a Gaussian variation of the height as implemented for XRR by the N\'{e}vot-Croce model \cite{nevot}.
The applied models consist of a homogenous Fe$_3$O$_4$/NiO bilayer on top of the respective substrate. 
This structural model involving the number of layers coincides with the layer model and the film thicknesses obtained from XRR calculations.
The obtained vertical layer distances ($c_{\textrm{NiO}}$/2 for NiO and $c_{\textrm{Fe}_3\textrm{O}_4}$/4 for Fe$_3$O$_4$) are shown in Fig.~\ref{XRD}(c). 


The dashed lines mark the bulk values of the magnetite and nickel oxide. 
Due to a larger unit cell of MgO(001) pseudomorphic growth of NiO on MgO results in an expansion of the NiO unit cell in lateral direction and, thus, a vertical compression and consequently a smaller vertical lattice distance. 
In the case of NiO grown on SrTiO$_3$(001) exactly the opposite is expected due to a smaller bulk unit cell of SrTiO$_3$ compared to NiO.
Thus, the vertical unit cell of NiO is larger than the bulk value as observed in the experiment. 

For the NiO layers on MgO the vertical layer distance exhibits a compressive strain (2.078\,\AA) and shows no dependence on the NiO thickness in the investigated range (cf.~Fig.~\ref{XRD}(c)). 
In the case of bilayers grown on SrTiO$_3$ the vertical lattice distance of NiO (2.095\,\AA) points to tensile strain as a result of the lateral compression. 
Further, there is no dependence on the NiO thickness, as well.

However, the situation is different for the relaxation of the magnetite films. 
Due to pseudomorphic growth of NiO on MgO the vertical layer distance of Fe$_3$O$_4$ grown on top of NiO/MgO is also slightly compressively strained but relaxes to higher values with increasing magnetite thickness. 
Its value relaxes from 2.0795\,\AA~for the 6.1\,nm thick magnetite film to 2.0885\,\AA~for the thickest magnetite film.
A strong relaxation with increasing film thickness of the magnetite can also be seen for magnetite films grown on NiO/SrTiO$_3$. 
The vertical lattice distance of Fe$_3$O$_4$ on NiO/SrTiO$_3$ is exposed to heavy tensile strain and decreases rapidly from 2.117\,\AA~for the thinnest film to 2.106\,\AA~for the 20.7\,nm thick magnetite film.

\begin{figure}[t]
\centering
\includegraphics[width=0.43\textwidth]{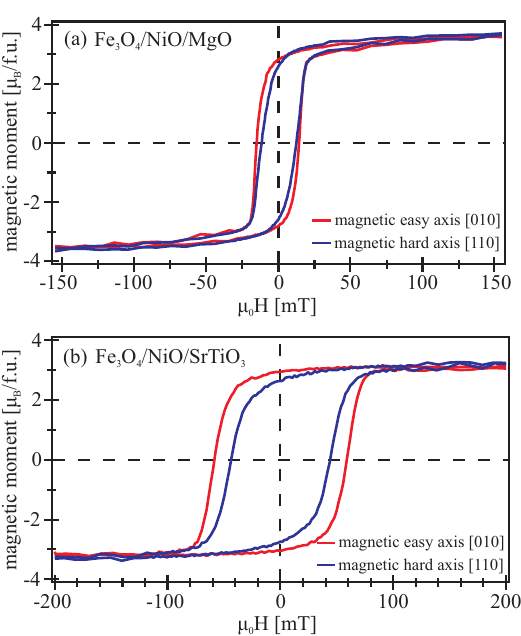}
\caption{VSM magnetization curves of magnetic easy and hard direction for (a) 21.5\,nm thick Fe$_3$O$_4$ film on NiO/MgO and (b) 20.7\,nm thick Fe$_3$O$_4$ film on NiO/SrTiO$_3$.}
\label{loops}
\end{figure}

\subsection{VSM}
Exemplary, the magnetic properties of the two thickest magnetite films on NiO/MgO and NiO/SrTiO$_3$ were studied by means of VSM.
The magnetization curves were measured for different azimuthal sample directions $\alpha$ between the substrate [100] direction and the applied magnetic field.
Figures~\ref{loops}(a) and \ref{loops}(b) show the magnetic moment per f.u. (formula unit) as a function of the magnetic field for the bilayers on MgO and SrTiO$_3$, respectively, for two different directions of the external magnetic field. 
For both samples a typical ferro(i)magnetic behavior can be observed. 
Here, the red curves recorded with the magnetic field applied in [010] direction of the substrates represent magnetic easy axes with a high magnetic remanence and coercive fields.
The blue curves recorded with the magnetic field applied in [110] direction exhibit the magnetic behavior of a magnetic hard axis due to a lower strength of the coercive field and a smaller magnetic remanence.
However, from magnetic saturation to magnetic remanence both investigated samples are not in a monodomain state, i.e. the squareness is not one in magnetic easy direction.
 
The Fe$_3$O$_4$ film on NiO/SrTiO$_3$ shows an enhanced coercive field compared to the magnetite film grown on NiO/MgO. 
One possible reason could be a higher density of grain boundaries due to the relaxation process, which supports pinned multidomain states that need larger magnetic fields to be switched. 
This is consistent with the weaker structural quality, e.g., high roughness, broad diffraction peaks, seen in the LEED, XRR and XRD measurements. 
Further, the saturation magnetization of the Fe$_3$O$_4$ film grown on NiO/MgO amounts to (3.7$\pm$0.3)\,$\mu_B$/f.u. and is rather close to the literature value of 4.07\,$\mu_B$/f.u. \cite{MagMom, MagMom2}.
In contrast, magnetite on NiO/SrTiO$_3$ shows a lower magnetic moment of (3.3$\pm$0.3)\,$\mu_B$/f.u., which may result from the antiferromagnetic coupling in the vicinity of anti-phase domain boundaries (APBs) \cite{APB}.

\begin{figure}[t]
\centering
\includegraphics[width=0.43\textwidth]{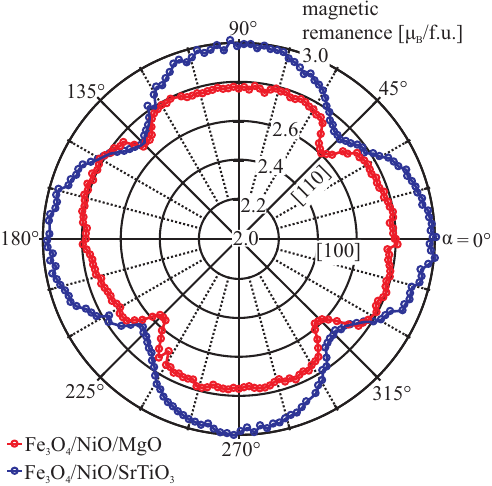}
\caption{Polar plot of the magnetic remanence depending on the azimuthal sample angle $\alpha$ of a 21.5\,nm thick Fe$_3$O$_4$ film on NiO/MgO (red) and 20.7\,nm thick Fe$_3$O$_4$ film on NiO/SrTiO$_3$ (blue).}
\label{remanence}
\end{figure}

The remanent magnetization as a function of azimuthal sample angle $\alpha$ is shown in Fig.~\ref{remanence} for both investigated samples.
The maxima of the magnetic remanence point into $\left\langle100\right\rangle$ directions for both Fe$_3$O$_4$ films on NiO/MgO and NiO/SrTiO$_3$ indicating the magnetic easy directions. 
Consequently, the magnetic hard axes are located in $\left\langle110\right\rangle$ directions.

\section{Discussion}

XPS measurements taken directly after deposition reveal stoichiometric Fe$_3$O$_4$ and NiO on both substrates independent of the film thicknesses. 
Due to the limited mean free path of the electrons only the near surface region ($\sim$5\,nm) of the layers could be characterized. 
In this region no evidence for the formation of non stoichiometric magnetite was observed. 
Pilard \textit{et al.} found a 1.5\,nm thick NiFe$_2$O$_4$ interfacial layer after depositing NiO above 610\,$^\circ$C on Fe$_3$O$_4$ \cite{MagPropBilayer}. 
Within the XPS measurements presented here the interfacial region could be detected only for the thinnest magnetite films showing spectral shape and binding energies typical for Ni$^{2+}$ in NiO stoichiometry.
Thus, there is no evidence for the formation of NiFe$_2$O$_4$ due to the lower growth temperature. 

Hard x-ray photoelectron spectroscopy (HAXPES) and x-ray magnetic circular dichroism (XMCD) measurements \cite{Kupper} of the same samples recorded after transport under ambient conditions show small traces of Fe$^{3+}$ excess on the surface of the bilayers grown on SrTiO$_3$. 
However, in deeper layers and at the interface the presence of stoichiometric NiO and Fe$_3$O$_4$ was confirmed excluding the formation of NiFe$_2$O$_4$ clusters or any interfacial layer also for thicker Fe$_3$O$_4$ films \cite{Kupper}. 
Consequently, very thin magnetite films tend to form maghemite at the surface after exposure to ambient air whereby thicker films seem to be more stable as it was reported before by Fleischer \textit{el al.} \cite{Fleischer}. 
Since \textit{in situ} XPS and LEED measurements taken after preparation under UHV conditions show no evidence for maghemite, a capping layer deposited directly after growth could prevent the possible oxidation process in the upper layers.

\textit{In situ} LEED measurements also verified the Fe$_3$O$_4$ stoichiometry of the iron oxide film showing the typical ($\sqrt{2}\times\sqrt{2})R45^{\circ}$ superstructure of the magnetite surface for all investigated films. 
Further, NiO films on both substrates exhibit the expected ($1\times1$) pattern due to the rock salt crystal structure. 
The diffraction spots of the magnetite and NiO films grown on SrTiO$_3$ are slightly broadened compared to the films grown on MgO indicating the formation of more surface defects due to the high lattice misfit.
Surface roughnesses obtained from the XRR analysis exhibit higher values for all films grown on SrTiO$_3$. 
While the roughness of the nickel oxide films deposited on SrTiO$_3$ is only about 0.5\,\AA\ higher than after deposition on MgO, the magnetite films on NiO/SrTiO$_3$ show initially almost doubled values compared to the magnetite films on NiO/MgO.
This result is consistent with the higher value for the defect density of Fe$_3$O$_4$/NiO/SrTiO$_3$ obtained from the LEED pattern analysis. 
Nevertheless, the XRR measurements provide distinct intensity oscillations indicating double layer structures and homogenous film thicknesses. 
Thus, the two layers do not intermix during the deposition process. 

The entire structure of the samples was investigated by XRD measurements of the specular CTR. 
For all samples the thickness determined by XRR agrees well with the number of layers obtained from XRD analysis where distinct Laue oscillations are observed.
The strong intensity oscillations reveal crystalline and well-ordered nickel oxide and magnetite films with homogeneous thicknesses on both substrates.

\begin{figure}[t]
	\centering
	\includegraphics[width=0.40\textwidth]{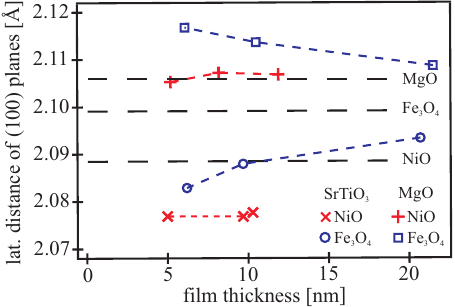}
	\caption{Lateral distance of (100) planes of all prepared magnetite and nickel oxide films calculated using Eq.~(\ref{eq:1}) and the vertical layer distances obtained from XRD analysis. The dashed lines denote the fully relaxed bulk values of MgO, Fe$_3$O$_4$ and NiO.}
	\label{latdist}
\end{figure}

The vertical layer distances of all NiO films show no dependence on the thickness in the investigated range. 
However, NiO and Fe$_3$O$_4$ films grown on MgO exhibit a vertical compressive strain while NiO and Fe$_3$O$_4$ films on SrTiO$_3$ show vertical tensile strain due to lattice matching at the interface. 
Based on elastic theory for continuum the vertical lattice constant $c$ for homogenous tetragonally (in-plane) distorted films is related to the lateral lattice constant $a$ via \cite{HashPoisson}
\begin{eqnarray}
\frac{\Delta c}{c} = \frac{2\nu}{\nu-1}\frac{\Delta a}{a}\,\,.
\label{eq:1}
\end{eqnarray}
For the calculation of the vertical layer distance for a completely strained film $\Delta a$ from pseudomorphic growth was used.
Assuming a Poisson number of $\nu$\,=\,0.21 for NiO \cite{NiOPoisson} the vertical layer distance of pseudomorphic nickel oxide on MgO was calculated to be 2.079\,\AA, hence, the NiO films grown on MgO are fully strained as clarified by Fig.~\ref{latdist} where lateral distances of (100) planes are presented. 

Above a critical thickness $d_c$ this strain should reduce rapidly due to the stable formation of dislocations. 
Following the model of Matthews and Blakeslee \cite{critthick}, the critical thickness $d_c$, at which the generation of misfit dislocation will begin, can be calculated by the formula 
\begin{eqnarray}
\frac{d_c}{b} = \frac{\left(1-\nu\,\cos^2\alpha\right)\,\left(\ln\left(\frac{d_c}{b}\right)+1\right)}{2\,\pi\,f\,(1+\nu)\,\cos(\lambda)}\,\,.
\label{eq:2}
\end{eqnarray}
Here, $b$ is the magnitude of the Burgers vector, $f$ the lattice mismatch, $\nu$ the Poisson ratio, $\alpha$\,=\,90$^\circ$ is the angle between the Burgers vector and the dislocation line and $\lambda$\,=\,45$^\circ$ is the angle between the Burgers vector and the direction both normal to the dislocation line and within the plane of the interface. 
For NiO films on MgO(001) the critical thickness is determined to 39\,nm. 
Since the studied films are below the critical thickness the absence of strain relaxation is in good agreement with this model. 
Similar results are also observed by Schemme \textit{et al.} \cite{SchemmeBilayer} for NiO films of different thicknesses up to 34\,nm grown on MgO(001). 
The experimental data of James \textit{et al.} \cite{NiOPoisson} show a strain relaxation above $\sim$40\,nm which is consistent with our observations and confirms Eq.~(\ref{eq:2}). 

Despite, the large misfit of -6.9\% between NiO and SrTiO$_3$ the XRD curves of all studied films also feature distinct Laue oscillations pointing to a good crystalline ordering. 
Assuming a complete lattice matching at the interface we calculate a vertical lattice distance of 2.161\,\AA\ for fully strained NiO films on SrTiO$_3$ (Eq.~(\ref{eq:1})) while we observe a film thickness independent value of 2.095\AA. 
The resulting lateral distances of the (100) planes calculated by Eq.~(\ref{eq:1}) of all investigated nickel oxide and magnetite films are presented in Fig.~\ref{latdist}. 
Thus, for the NiO films grown on SrTiO$_3$ the remaining lateral strain only amounts to -0.6\% (cf.~Fig.~\ref{latdist}). For the critical thickness Eq.~(\ref{eq:2}) reveals a value of 3.5\,nm.
All prepared NiO films are well above the critical thickness. 
Thus, the observed strong strain relaxation seems to be reasonable although they are not completely relaxed. 
We assume that the residual strain cannot be removed from the film due to kinetic barriers preventing the film to relax completely. 
Similar strain behavior was reported by Zhang \textit{et al.} for NiO films of 2\,nm thickness grown by pulsed laser deposition on SrTiO$_3$ substrates. 
In contrast to our findings, a complete relaxation for NiO films of thicknesses above 10\,nm was observed, probably driven by higher deposition temperature \cite{Zhang2017}.

In the case of Fe$_3$O$_4$ on NiO/MgO, we calculate a vertical layer distance for a fully strained film of 2.092\,\AA~and a critical thickness of 105\,nm ($\nu$\,=\,0.356 \cite{PoissonMagn}, $f$\,=\,0.3\%) applying Eq.~(\ref{eq:1}) and Eq.~(\ref{eq:2}), respectively. 
Here, the misfit $f$ coincides with the misfit of magnetite on MgO since the growth of NiO on MgO is pseudomorph adapting its lateral lattice constant (cf.~Fig.~\ref{latdist}).  
All our investigated magnetite films on NiO/MgO are strongly strained having a lower vertical layer distance than received by Eq.~(\ref{eq:1}).  
Futher, the calculated lateral layer distance of all prepared Fe$_3$O$_4$ films is larger than of the NiO films pseudomorphically grown on MgO (cf.~Fig.~\ref{latdist}).
Consequently, the magnetite films are exposed to much higher tensile lateral strain as expected from classical growth theory. 
This effect may be attributed to the unpreventable formation of APBs which is not considered in the simple theories of epitaxial growth and relaxation via misfit dislocations. 
Thus, we assume that APBs expose additional tensile strain to the magnetite film. 
This result is in contrast to the compressive strain due to APBs as reported for magnetite films (thickness range 85\,-\,600\,nm) directly grown on MgO(001) by magnetron sputtering \cite{Balakrishnan04}.
As shown in Fig.~\ref{XRD}(c) the measured vertical layer distance approaches the bulk value with increasing Fe$_3$O$_4$ thickness. 
However, the bulk value is not reached even for the thickest magnetite film on NiO/MgO studied here. 
On one hand, this effect is very surprising since the predicted critical thickness of 105\,nm is beyond the thicknesses under consideration here. 
On the other hand, this behavior of partial relaxation below the calculated critical thickness also coincides with the results reported by Schemme et al. \cite{SchemmeBilayer}.
We also attribute this effect to the unpreventable formation of APBs which is not considered in the simple theories for the nucleation of misfit dislocations. 
Thus, APBs seem to lower the kinetic barrier for the formation of dislocations.

Regardless of the low remaining compressive strain between the Fe$_3$O$_4$ and NiO/SrTiO$_3$, these magnetite films are less structurally ordered than the magnetite films grown on NiO/MgO.
While the crystalline quality of the NiO films on SrTiO$_3$ is constantly high independent of the film thickness, the strength of Laue oscillations of the Fe$_3$O$_4$ films grown on top of NiO/SrTiO$_3$ decreases with increasing magnetite thickness. 
This result is supported by the high surface roughness of the magnetite films obtained from the XRR measurements as well as by the broadened diffraction spots seen in the LEED pattern. 

Assuming pseudomorphic growth of magnetite on the strained NiO film with remaining lattice mismatch of -1\%, the vertical layer distance of a fully strained magnetite film is calculated to 2.123\,\AA\ using Eq.~(\ref{eq:1}). 
The measured value of 2.117\,\AA\ for the 5\,nm thick magnetite film is already lower than the expected value for pseudomorphic growth (cf.~Fig.\ref{XRD}(c)). 
Thus, this magnetite film is already partially relaxed and shows vertical and lateral strain of 0.9\% and 0.8\%, respectively (cf.~Fig.~\ref{latdist}).
With increasing thickness of the magnetite film the vertical and lateral layer distances strongly relaxes further to 2.104\,\AA\ and 2.093\,\AA, respectively, for the 20.7\,nm thick film.
Again this effect contradicts classical relaxation theory via dislocation formation from which the critical film thickness of 27\,nm is obtained using Eq.~(\ref{eq:2}). 
Consequently, the formation of grain boundaries and structural defects, e.g., APBs during the initial stage of film growth may support the formation of misfit dislocations and, thus, a faster relaxation process. 
In addition, as stated above, the lateral tensile strain due to APBs may cause a larger lateral layer distance compared to pseudomorphic growth on the strained NiO film.

VSM measurements of the two thickest magnetite films on NiO/MgO and NiO/SrTiO$_3$ reveal ferro(i)magnetic behavior for both samples. 
However, the Fe$_3$O$_4$ film grown on NiO/SrTiO$_3$ shows enhanced coercive field compared to the film on NiO/MgO possibly caused by a higher density of grain boundaries and, thus, a formation of more pinning centers as confirmed by the LEED analysis. 
This behavior coincides with the weaker structural ordering and a higher surface roughness of the magnetite films on NiO/SrTiO$_3$ also seen in the XRD and XRR measurements. 
An increased coercive field for magnetite films grown on SrTiO$_3$ caused by a higher surface roughness or strain have also been reported in Refs. \cite{MagAnisotrSTO,MagAnisotr2}.

The obtained saturation magnetization values of Fe$_3$O$_4$ grown on NiO/MgO and NiO/SrTiO$_3$ coincide within the error tolerances with the values determined by XMCD \cite{Kupper}.
Additionally, the value of Fe$_3$O$_4$ film on NiO/MgO is also rather close to the ideal theoretical value as well as to the experimental bulk moment of magnetite of 4.07$\mu_B$/f.u. \cite{MagMom,MagMom2,magneticmomentmagnetite}, whereas Fe$_3$O$_4$ on NiO/SrTiO$_3$ exhibits a lower value. 
A reduced magnetic moment has also been reported for Fe$_3$O$_4$/SrTiO$_3$ systems possibly caused by a large density of APBs induced by high lattice mismatch \cite{MagSTO,Mag111STO}.
This result is supported by a weaker structural ordering as well as higher coercive fields and, thus, a higher density of grain boundaries observed for Fe$_3$O$_4$ on NiO/SrTiO$_3$.

Further, both investigated samples show a fourfold magnetic in-plane anisotropy with magnetic easy axes aligned along the $\left\langle100\right\rangle$ directions.
For thin magnetite films on MgO(001) the magnetic easy axes are mostly reported to point into $\left\langle110\right\rangle$ directions \cite{MagAnisotr2,MagAnisotrMgO,Schemme3} as expected from bulk properties of Fe$_3$O$_4$.
But also a magnetic isotropic behavior \cite{Schemme3,kale2001} or magnetic easy axes aligned in $\left\langle100\right\rangle$ directions \cite{prieto2015} are presented in literature for Fe$_3$O$_4$/MgO(001).
Moreover, magnetite films grown on an iron buffer layer deposited on MgO(001) exhibit a magnetic in-plane anisotropy with magnetic easy axes parallel to $\left\langle100\right\rangle$, too \cite{SchemmeFeBuffer}. 
For Fe$_3$O$_4$ films on SrTiO$_3$(001) also different orientations of the magnetic easy axes are reported. 
While Kale et al. observed a fourfold magnetic anisotropy with magnetic easy axes pointing into $\left\langle110\right\rangle$ directions \cite{kale2001}, magnetic easy axes aligned along the $\left\langle100\right\rangle$ directions are presented in Refs. \cite{MagAxisMagSTO,prieto2015}.
All these observations show that the magnetic properties of magnetite are highly affected by the interface between the film and substrate and can be influenced by the deposition conditions, lattice mismatch or stoichiometric deviations. 
In addition, we assume that a tetragonal distortion of the films can influence the spin-orbit coupling which may lead to modified magnetocrystalline anisotropy constants \cite{spinorbitanis} and, thus, altered directions of magnetic easy and hard axes. 

\section{Summary}

We present a comparative study on the structural and magnetic properties of Fe$_3$O$_4$/NiO bilayers grown on MgO(001) and Nb-doped SrTiO$_3$(001). 
Stoichiometric magnetite and NiO films with homogenous thicknesses were found on both substrates in the investigated thickness range (5-20\,nm).
Detailed analysis of the XRD measurements reveal a high crystallinity of the NiO films independent of the underlying substrate or film thickness. 
However, magnetite films grown on NiO/SrTiO$_3$ exhibit a weaker structural ordering and higher surface roughness compared to the films grown on NiO/MgO induced by a large lattice mismatch and the resulting relaxation process.
Further, the bilayers exhibit a vertical compressive strain on MgO but a tensile strain in vertical direction on SrTiO$_3$ as a result of lateral compression.
The weaker crystalline structure of Fe$_3$O$_4$ on NiO/SrTiO$_3$ affects the magnetic properties leading to an enhanced coercive field and a reduced magnetic moment compared to magnetite on NiO/MgO. 
Nevertheless, these Fe$_3$O$_4$/NiO bilayers on MgO and SrTiO$_3$ substrates are expected to show large thermoelectric effects based on the thermal generation of spin currents (spin Seebeck effect) \cite{ramos13,ramos16,SSE} supported by the antiferromagnetic NiO layer \cite{SpinCurrentAFM,SSEAFM}. 

Additionally, both systems show a fourfold magnetic in-plane anisotropy with magnetic easy axes pointing in $\left\langle100\right\rangle$ directions which are 45$^\circ$ rotated to the well known magnetic easy axes directions of thin magnetite films on MgO(001) as expected from bulk properties.
One potential reason may be a modified spin-orbit coupling as a result of the tetragonal distortion of the films leading to altered magnetocrystalline anisotropy. 
A detailed understanding of these bilayers is of utmost importance since they are excellent candidates for potential spintronic and spin caloritronic applications. 
Therefore, this behavior deserves further studies to shed more light on this interesting change of the magnetic anisotropy of Fe$_3$O$_4$ thin films grown on NiO/MgO(001) and NiO/SrTiO$_3$(001).\\

\section*{Acknowledgments}

The authors gratefully acknowledge the financial support by the Deutsche Forschungsgemeinschaft (DFG) via grants No. KU2321/2-1 and KU3271/1-1. Portions of this research were carried out at beamline I811, MaXLab synchrotron radiation source, Lund University, Sweden. Funding for the beamline I811 project was kindly provided by The Swedish Research Council and The Knut och Alice Wallenbergs Stiftelse. Additional experiments were performed at the X04SA beamline at the Swiss Light Source synchrotron radiation source at Paul Scherrer Institute, Villigen, Switzerland. We like to thank the I811 and X04SA beamline staff for experimental support.

\bibliography{Manuscript}

\end{document}